\documentclass[reprint,amsmath,amssymb,aps]{revtex4-1}

\usepackage{graphicx}
\usepackage{bm}
\usepackage{physics}
\usepackage{natbib}
\usepackage{amsfonts,amsthm,bbold}
\usepackage{tikz}
\usepackage{mathtools}
\usepackage[justification=centerlast]{caption}

\def\at{
  \left.
  \vphantom{\int}
  \right|
}

\begin{document}

\preprint{APS/123-QED}

\title{Statistical analysis of quantum entangled network generation} 

\author{Scott E. Vinay}
 \email{svinay1@sheffield.ac.uk}
\author{Pieter Kok}
 %\email{p.kok@sheffield.ac.uk}
\affiliation{ Department of Physics and Astronomy, University of Sheffield}

\date{\today}

\begin{abstract}
We develop techniques to analyse the statistics of completion times of non-deterministic elements in quantum entanglement generation, and how they affect the overall performance as measured by the secret key rate. By considering such processes as Markov chains, we show how to obtain exact expressions for the probability distributions over the number of errors that a network acquires, as well as the distribution of entanglement establishment times. We show how results from complex analysis can be used to analyse Markov matrices to extract information with a lower computational complexity than previous methods. We apply these techniques to the Innsbruck quantum repeater protocol, and find that consideration of the effect of statistical fluctuations tightens bounds on the secret key rate by 3 orders of magnitude. We also use the theory of order statistics to derive tighter bounds on the minimum quantum memory lifetimes that are required in order to communicate securely.
\end{abstract}
\maketitle

\section{Introduction} \label{sec:intro}

The ability to construct large-scale quantum networks between two or more parties is a necessary precursor to the general deployment of entanglement-based quantum key distribution as a ubiquitous alternative to classical encryption \cite{ekert1991quantum,vazirani2014fully}, as well as the creation of measurement-based quantum computers \cite{briegel2009measurement}. Implementations of such networks would range from Bell states for point-to-point communication over large distances \cite{bell2001einstein,Briegel1998QuantumCommunication}, to highly connected cluster states \cite{Kok2010IntroductionProcessing} and a complete distributed quantum Internet \cite{kimble2008quantum}. Many theoretical proposals have been put forward for different schemes to implement these tasks, and in general the construction of these quantum networks requires the use of probabilistic elements. For example, many probabilistic methods for the generation of entanglement between nodes of a network have been proposed \cite{Barrett2005EfficientOptics,Campbell2008Measurement-basedLoss,Benjamin2006BrokeredComputation,Childress2006Fault-tolerantEmitters}, as well as many high-level schemes that take advantage of such methods, such as entanglement-based quantum repeaters \cite{Pant2016Rate-distanceRepeaters,Barrett2010ScalableEnsembles,Matsuzaki2010ProbabilisticAccumulation,Vinay2017PracticalCommunication,Azuma2015All-photonicRepeaters,Duan2001Long-distanceOptics.} and entanglement distillation \cite{Deutsch1996QuantumChannels,Dur2007EntanglementCorrection}. Probabilistic methods are also used in the implementation of non-linear unitary operations on optical states, such as those used in linear optical quantum computation \cite{knill2001scheme,Browne2005Resource-efficientComputation,Kok2007LinearQubits} and code-based repeaters \cite{Munro2012QuantumMemories,Ralph2005Loss-TolerantQubits}, as well as schemes for making measurements of states in a way that is protected against particle loss \cite{Varnava2006LossCorrection}. The presence of such probabilistic components means that a complex composite protocol will likely take many attempts before completing its task. When a single element fails, this could result in the entire process, or a subsection of it, needing to be restarted. It may also result in \emph{waiting errors}. This is where one part of the protocol finishes, but accumulates errors while waiting for another part to complete.

Typically, in many analyses of quantum network systems, the full depth of statistical information that may be gleaned from the full probability distributions over completion times or error distributions is neglected in favor of a simpler analysis, such as analysing the average values. However, this can result in too limited a characterization of the protocol, and one that may miss essential features. For example, a situation that is commonly considered in the context of quantum communication is the time taken to generate a set of entangled states between Alice and Bob, which may be distilled in order to generate a smaller number of higher-fidelity pairs. If we consider that all pairs connect after some average time, $t$, then the secret key rate will scale linearly with the number of states that we are trying to connect in parallel, and inversely proportional with $t$. In reality, not all pairs of entangled states will establish at the same time. However, if we intend to use all of them for distillation, then the pairs that establish first will have to be stored on quantum memories, and the fidelities of these states will decay while they wait for the other pairs to complete. It will therefore not necessarily be advantageous to have a greater number of pairs try to establish their entanglement in parallel. A good understanding of the distribution of times taken by a protocol and the error probabilities is thus essential for any analysis of a protocol.

We begin with some general methods that may be used for the analysis of probabilistic processes using Markov chain analysis. Markov chains have recently been applied to quantum networks by Shchukin, Schmidt and van Loock \cite{shchukin2017waiting}. We build upon these techniques in order to include errors in a natural way, as well as introducing new analytic techniques to greatly reduce the computational burden that comes with any deep analysis of Markov chains. In Section \ref{sec:markov} we explain how one may construct Markov matrices for probabilistic processes, and how the matrices for larger compound processes can be constructed from the matrices of smaller processes. We show also how we can find $\{p_t|\hspace{.5mm} t\in\mathbb{N}\}$ from such matrices, where $p_t$ is the probability that the process will complete at time $t$. In Section \ref{sec:cauchy} we show how one may find the probability-generating function (PGF) from the Markov matrix. We then show how one may solve the PGF to find the completion time distribution such that the computational complexity of finding $p_t$ is decreased by a factor of $n$ compared to using the matrix alone, where $n$ is the dimension of the matrix.

In Section \ref{sec:fft} we show how to calculate the probability distribution for the number of times that a given event in a process occurs. This rather general method may be used to calculate the distribution of the number of errors that will accumulate in the running of a process, both on average and conditioned on the completion time.

In Section \ref{sec:innsbruck} we examine a modification of the Innsbruck protocol for distillation-based quantum repeaters \cite{Briegel1998}, where the available quantum memories at a repeater station are bunched. By this, we mean we separate the available pairs of quantum memories between each pair of repeater stations into bunches of fixed size which are then distilled once all entanglement connections within the bunch have completed. We apply the techniques developed here to estimate the best values for the sizes of these bunches. This allows for a richer characterization of the secret key rates reachable by a protocol than may be learned from an analysis that does not account for the statistical factors that are captured by the Markov chain formulation. Finally, in Section \ref{sec:akrb} we consider a simplification of our statistical analysis of the Innsbruck protocol. By considering bounds on the order statistics of completion times of certain elements within the protocol, we derive bounds on the secret key rates. This allows us to identify minimum experimental parameters that must be reached in order to securely communicate over a repeater network of many sections.

\section{Markov chains} \label{sec:markov}

\begin{figure}[t]
   \centering
     \includegraphics[width=.5\columnwidth]{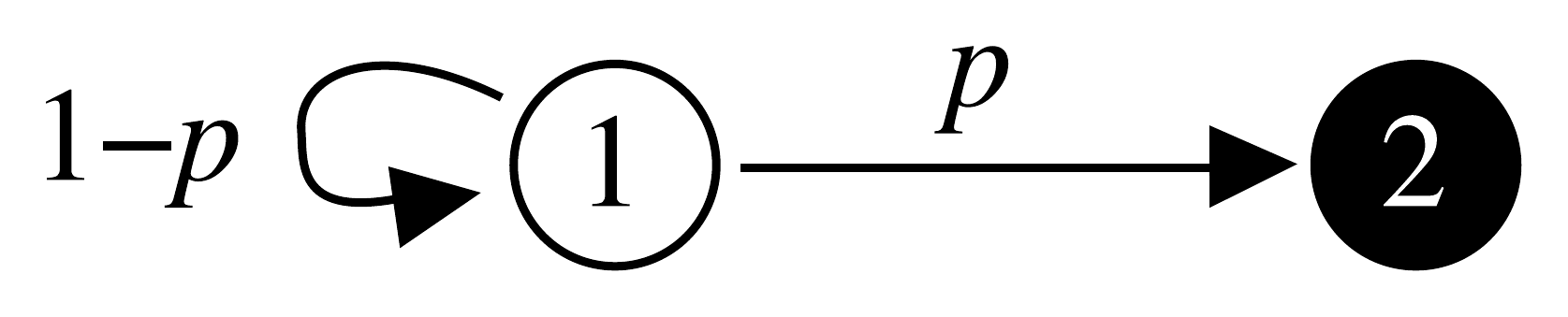}
     \captionsetup{justification=justified}
     \caption{Graph for simple probabilistic entanglement generation. Each node represents a state that the system may be in at any one time. Transitions between nodes are considered to all take the same length of time, and occur with probabilities indicated by the weight of the edge. Terminating node shown in black.}
     \label{fig:GP}
\end{figure}

Let $\mathcal{P}$ be some process that may be decomposed into events taking place across a series of discrete time-steps. This process may be summarised by a directed graph, $G_\mathcal{P}$, which is a flowchart showing possible paths of progression. Each node represents a unique state that the process may be in at any one time. The edges leading away from each node are the possible events (with the traversal of an edge being considered to take one time-step), with the weight of each edge representing the probability that that step will be taken. Each graph must include at least one terminating node (with no edges leaving it) representing the termination of the process. For example, if $\mathcal{P}$ is the protocol of establishing entanglement between a single pair of quantum memories by a probabilistic process that succeeds with probability $p$, then $G_\mathcal{P}$ is given by Fig.~\ref{fig:GP} (see \footnote{Here we have used the convention that an absorbing node of the process has no edges leading away from it. Many standard texts on Markov chains use the convention that absorbing nodes should transition to themselves with probability 1.}). Since $\mathcal{P}$ is probabilistic, the time that it takes to complete is represented by a random variable, $T$, that takes on value $t$ with probability $p_t$.

From here, we can form the square Markov matrix for the process, $M_\mathcal{P}$, which is the adjacency matrix of $G_\mathcal{P}$ \footnote{These Markov matrices are such that the columns sum to either 0 (for terminating nodes) or 1.}. That is to say, $M_\mathcal{P}\at_{i,j}$ is the weight of the edge of $G_\mathcal{P}$ leading from node $j$ to node $i$ for some fixed labeling of $G_\mathcal{P}$. This immediately gives us an operational method to find $\{p_t\}$: if we let $I_\mathcal{P}$ be the set of indices for terminating nodes, then we may say that

\begin{equation} \label{eq:pt_mat_mult}
p_t = \sum_{i\in I_\mathcal{P}} M_\mathcal{P}^t\at_{1,i},
\end{equation}

\noindent where the node representing the start of the process is given the label 1.

From the matrices for simple processes we can build up matrices for more complex processes. Consider two processes $\mathcal{P}_1$ and $\mathcal{P}_2$. We wish to concatenate these to form the process $\mathcal{P}_3$, which consists of $\mathcal{P}_1$ and $\mathcal{P}_2$ being run simultaneously but independently. For each unique pair of states with one chosen from $\mathcal{P}_1$ and one chosen from $\mathcal{P}_2$, we should assign a unique state in $\mathcal{P}_3$. Additionally, for two such pairs of states, $s_{\mathcal{P}_1}^1, s_{\mathcal{P}_2}^1$ and $s_{\mathcal{P}_1}^2, s_{\mathcal{P}_2}^2$, then the independence of $\mathcal{P}_1$ and $\mathcal{P}_1$ implies that the probability to move from the state representing $(s_{\mathcal{P}_1}^1, s_{\mathcal{P}_2}^1)$ to $(s_{\mathcal{P}_1}^2, s_{\mathcal{P}_2}^2)$ in $\mathcal{P}_3$ should be given by ${p(s_{\mathcal{P}_1}^1 \rightarrow s_{\mathcal{P}_1}^2) \cdot p(s_{\mathcal{P}_2}^1 \rightarrow s_{\mathcal{P}_2}^2)}$. Therefore if we consider $\mathcal{P}_3$ to have finished when $\mathcal{P}_1$ \emph{or} $\mathcal{P}_2$ have finished, then

\begin{equation} \label{eq:mP_tensor_or}
M_{\mathcal{P}_3} = M_{\mathcal{P}_1} \otimes M_{\mathcal{P}_2}.
\end{equation}

We may instead wish to wait until both $\mathcal{P}_1$ \emph{and} $\mathcal{P}_2$ have completed before considering $\mathcal{P}_3$ to have completed. In this case we should add an element to the matrix for each subprocess that keeps the system on that terminating node until the other subprocess has completed. The composite matrix is therefore

\begin{equation} \label{eq:mP_tensor_and}
\begin{split}
M_{\mathcal{P}_3} = &\left[M_{\mathcal{P}_1}+\textrm{diag}(\mathbf{I}_{\mathcal{P}_1})\right] \otimes \left[M_{\mathcal{P}_2}+\textrm{diag}(\mathbf{I}_{\mathcal{P}_2})\right] \\&- \textrm{diag}(\mathbf{I}_{\mathcal{P}_1}\otimes \mathbf{I}_{\mathcal{P}_2}).
\end{split}
\end{equation}

\noindent where $\mathbf{I}_{\mathcal{P}}\!\at_i = 1$ if $i\in I_\mathcal{P}$, and 0 otherwise, and $\textrm{diag}(\mathbf{I})$ is a matrix with the elements of $\mathbf{I}$ on the diagonal, and with zeros elsewhere. 

Suppose that instead we consider $\mathcal{P}_3$ to consist of $\mathcal{P}_1$ followed by $\mathcal{P}_2$. When we reach the terminating nodes of $\mathcal{P}_1$, the next time-step will have us arrive at the first node of $\mathcal{P}_2$. Then

\begin{equation} \label{eq:oplus_mat}
\left[M_{\mathcal{P}_3}\right]_{i,j} = \left[M_{\mathcal{P}_1}\oplus M_{\mathcal{P}_2}\right]_{i,j} + \sum_{k \in T_{\mathcal{P}_1}}\delta_{i,k}\delta_{j,n_{\mathcal{P}_1}+1} %XXChange this
\end{equation}

\noindent where $n_{\mathcal{P}_1}$ is the number of nodes in $G_{\mathcal{P}_1}$ or the number of rows or columns in $M_{\mathcal{P}_1}$).

%XX change this
It may also be the case that different parts of a process take different lengths of time, instead of the above construction which assumes that each event takes a single time-step. Suppose that within some process, $\mathcal{P}$, we have some events (edges on $G_\mathcal{P}$) that take some time $k_1$, and some that take $k_2$, where $k_2 \geq k_1$. We can decompose $M_\mathcal{P}$ as $M_{\mathcal{P},k_1} + M_{\mathcal{P},k_2}$, such that all elements in $M_{\mathcal{P},k_1}$ represent events that take $k_1$, and similar for $M_{\mathcal{P},k_2}$.
%Remove space
From this we can create a new process matrix $M_\mathcal{P}^\prime$ which properly accounts for the fact that events in subprocess $P_1$ can be done many times for each time that $P_2$ can be done. This is given by 

\begin{equation} \label{eq:unequal_timings}
\begin{split}
&\left[M_\mathcal{P}^\prime\right]_{i,j} = \sum_k \Bigg\{\left[M_{P,k_1}^{\lceil k_2/k_1 \rceil}\right]_{i,k} + \\ & \left(1-\sum_i\left[M_{P,k_1}^{ \lceil k_2/k_1 \rceil}\right]_{i,k}\right)\delta_{i,k}  \Bigg\}
\left[M_{P,k_2}\right]_{k,j}
\end{split}
\end{equation}

As such, $p_t$ calculated from $M_\mathcal{P}^{\prime t}$ will represent the probability that the process completes after $t$ applications of $P_2$ and $t k_2/k_1$ applications of $P_1$. It should be noted that the modification of process matrices to account for timing differences should be done \emph{before} creating composite matrices by tensor products.

Using Markov matrices along with Eq.~(\ref{eq:pt_mat_mult}) is a simple way to calculate the completion times of a process, although it is not necessarily the most efficient. Multiplying $M_\mathcal{P}$ by itself takes $n_\mathcal{P}^2$ elementary multiplications. Given some algorithm for calculating exponentials that has a number of operations that scales asymptotically as $f_\textrm{exp}(t)$ for the calculation of $k^t$ for some constant $k$, we find that the calculation of $p_t$ scales asymptotically as 

\begin{equation} \label{eq:order_mag_mat_mult}
\mathcal{O}\!\left(p_t\hspace{1mm} \textrm{by matrix mult}\right) = \mathcal{O}\!\left(n_\mathcal{P}^2 f_\textrm{exp}(t)\right).
\end{equation}

In the next section we derive a method by which this may be reduced by a factor of $n_\mathcal{P}$.

\section{Probability generating functions} \label{sec:cauchy}

In this section we show how an approach based on probability generating functions (PGFs) and complex analysis can lead to formulas for $p_t$ that are faster to compute than the matrix multiplications of Eq.~(\ref{eq:pt_mat_mult}). 

The probability generating function of a distribution $\{p_t\}$ corresponding to the completion times for a process $\mathcal{P}$ is defined as the polynomial

\begin{equation} \label{eq:pgf}
f_\mathcal{P}(z) = p_0 + p_1 z + p_2 z^2 + p_3 z^3 + \cdots,
\end{equation}

\noindent where $z$ is a complex variable, and $p_t$ are constants to be determined based on $\mathcal{P}$. Given the PGF associated with some process, the elements $p_t$ may be found by calculating the coefficients of the various terms by finding the derivatives: 

\begin{equation} \label{eq:eachDerivative}
p_t = \frac{1}{t!}\frac{\textrm{d}^t f_\mathcal{P}(z)}{\textrm{d}z^t}\at_{z=0}.
\end{equation}

In order to write down the PGF, it may seem like we need to already know all of $p_t$. However, we can calculate $f_\mathcal{P}(z)$ directly from $G_\mathcal{P}$. Consider a node in $G_\mathcal{P}$, $x$, with one edge leading to node $y$ with probability 1. Let $f_\mathcal{P}^{(x)}(z)$ be the PGF for the system when we start at node $x$. Since the system will take exactly one time-step longer to complete when we start at $x$ than when we start at $y$, we can say that $f_\mathcal{P}^{(x)}(z) = z f_\mathcal{P}^{(y)}(z)$. Now suppose that $x$ has two edges leading away from it to nodes $y_1$ and $y_2$ with probabilities $p(y_1)$ and $p(y_2)$ respectively. Then, $f_\mathcal{P}^{(x)}(z) = z\hspace{.5mm} p(y_1) f_\mathcal{P}^{(y_1)}(z) + z\hspace{.5mm} p(y_2) f_\mathcal{P}^{(y_2)}(z)$. By extension, we may say that

\begin{equation}\label{eq:graph_rules_2}
f_\mathcal{P}^{(j)}(z) = 
\begin{cases}
\sum_j \left[M_\mathcal{P}\right]_{i,j} z f_\mathcal{P}^{(i	)}(z) \hspace{2mm}\textrm{if}\hspace{2mm} i \not\in I_\mathcal{P}, \\
\makebox[0pt][l]{1}\phantom{\sum_j \left[M_\mathcal{P}\right]_{i,j} z f_\mathcal{P}^{(j)}(z) } \hspace{2mm}\textrm{if}\hspace{2mm} i \in I_\mathcal{P},
\end{cases}
\end{equation}

\noindent where the sum runs over all columns in the matrix, which is an eigenvalue equation. The PGF of the process as a whole $[f_\mathcal{P}(z)]$ may be identified with the PGF of the initial node $[f_\mathcal{P}^{(1)}(z)]$. In particular, $f_\mathcal{P}(z)$ is the first element of the eigenvector of $\tilde{M}_\mathcal{P}(z)$ with eigenvalue 1, normalised such that the $k^\textrm{th}$ element is 1 for any $k\in I_\mathcal{P}$, where

\begin{equation}
\tilde{M}_\mathcal{P}(z) = z M_\mathcal{P}^T + \textrm{diag}(\mathbf{I}_\mathcal{P}).
\end{equation}

However, a problem may arise in the process of finding the set of eigenvectors. We wish to retain $z$ as an open variable in the PGF, which means that many of the fast methods for finding eigenvalues of matrices cannot be used, since they rely on numerical methods. In order to find an eigenvector of a completely general matrix, $M$, we need to be able to solve the characteristic equation $\left|M-\lambda \mathbb{1}\right|=0$. This involves exactly solving a polynomial of order $n_\mathcal{P}$, which cannot in general be done for $n_\mathcal{P}\geq 5$. Instead, we use the fact that the eigenvalue is 1, so that $\tilde{M}\bm{f_\mathcal{P}}=\bm{f_\mathcal{P}}$, where $\bm{f_\mathcal{P}}$ is the vector with $i^\textrm{th}$ element equal to $f_\mathcal{P}^{(i)}$, and say that 

\begin{equation}
\begin{split}
&f_\mathcal{P}(z) = \left[\bm{f_\mathcal{P}}\right]_1/\left[\bm{f_\mathcal{P}}\right]_k\\
&\bm{f_\mathcal{P}} = \textrm{Null}\!\left[\tilde{M}_\mathcal{P}(z) - \mathbb{1}\right],
\end{split}
\end{equation}

\noindent for any $k\in I_\mathcal{P}$.
Note that we have used a slight abuse of notation and specified that $\bm{f_\mathcal{P}}$ is equal to the null space itself and not a particular vector in the null space. This is because the null space has a dimension of 1. We can see this by the fact that, if $M_\mathcal{P}$ is a Markov matrix, then $\tilde{M}_\mathcal{P}^T(z)$ must also be Markovian at $z=1$. Moreover, the sum of all values in each column of $\tilde{M}_\mathcal{P}^T(1)$ will equal 1, which means that $\tilde{M}_\mathcal{P}^T(z)$ fits the usual definition of a stochastic matrix found in standard Markov chain textbooks. All stochastic matrices have exactly one eigenvalue at 1 \cite{privault2013understanding}, and so the other eigenvalues of $\tilde{M}_\mathcal{P}(z)$ must either be never equal to 1 or $z$-dependent. 

Having found the PGF, we want to use it with Eq.~(\ref{eq:eachDerivative}) to find $\{p_t\}$. Manually calculating the first few derivatives of the PGF may be possible. However the task soon becomes difficult for higher-order terms. By using Cauchy's differential formula \cite{mathews2012complex}, we find not only an easy way to compute higher derivatives, but a closed-form expression for an \emph{arbitrary} derivative that can easily be calculated without needing to calculate all lower derivatives. The formula states that for some point $a \in \mathfrak{S}$, 

\begin{equation} \label{eq:Cauchy_derivative}
\frac{1}{t!}\frac{\textrm{d}^t f_\mathcal{P}(z)}{\textrm{d}z^t}\at_{z=a} = \frac{1}{2\pi i}\oint_{\partial \mathfrak{S}} \frac{f_\mathcal{P}(z)}{(z-a)^{t+1}}\textrm{d}z,
\end{equation}

\noindent where $\partial \mathfrak{S}$ is the boundary of $\mathfrak{S}$; a compact subset of $\mathbb{C}$ on which $f_\mathcal{P}(z)$ is analytic.  

Let $a=0$ and $g_t(z) = f_\mathcal{P}(z)/z^{t+1}$. We will evaluate the integral of $g_t(z)$ on a circle centered on $z=0$. If the contour encloses no poles except the one at 0 due to the $z^{-(t+1)}$ term, then this is equivalent to finding the residue of the pole of $g_t(z)$ at 0. Suppose that $f_\mathcal{P}(z)$ scales as $\mathcal{O}(z^t_0)$ as $\left|z\right|\rightarrow\infty$. Then the integral of $g_t(z)$ on a circular path of radius $R$ will tend to 0 as $R \rightarrow \infty$ for all $t>t_0$ (since $\textrm{d}z = \left|z\right| \textrm{d}\theta$). However, by Cauchy's residue theorem, this integral must \emph{also} be equal to the sum of all residues of $g_t(z)$ in $\mathfrak{S}$. This includes the pole at 0, which we get from the $z^{-(t+1)}$ term, and the poles elsewhere in the complex plane, which are the poles of $f_\mathcal{P}(z)$. Therefore the sum of the residues of all poles must be equal to 0 for $t>t_0$. The residue at $z=0$ cannot be easily directly calculated since it is a non-simple pole, but we can calculate it indirectly since we know it must be equal to the negative of the sum of the residues of the other poles, which are in general simple. We therefore arrive at the main result of this section:

\begin{equation} \label{eq:p_t_general}
p_t = -\!\sum_i \textrm{Res}\left[ \frac{f_\mathcal{P}(z)}{z^{t+1}} , z_i \in \mathbb{P}\!\left(f_\mathcal{P}\right) \right],
\end{equation}
\noindent where $\mathbb{P}\!\left(f\right)$ is the set of singularities of $f_\mathcal{P}(z)$.

As a corollary, we may use this method to easily find expectation values for completion times of such processes. Consider that 

\begin{equation} \label{eq:tavg}
\left\langle T \right\rangle = \sum_t t \hspace{1mm} p_t. 
\end{equation}

\noindent If we use the fact that 

\begin{equation}
\textrm{Res}\left[ \frac{f_\mathcal{P}(z)}{z^{t+1}} , z_i \right] = \frac{\textrm{Res}\left[ f_\mathcal{P}(z) , z_i \right]}{{z_i}^{t+1}},
\end{equation}

\noindent since $f_\mathcal{P}(z)$ has no pole at 0, we can write Eq.~(\ref{eq:tavg}) as 

\begin{equation}
\left\langle T \right\rangle =-\sum_i \textrm{Res}\left[ f_\mathcal{P}(z) , z_i \right] \sum_t \frac{t}{z_i^{t+1}},
\end{equation}

\noindent where the $z_i$ sum is implicitly over the poles of $f_\mathcal{P}(z)$. We can use the identity ${\sum_{n=0}^\infty n \hspace{0.5mm} x^{n-1} = (1-x)^{-2}}$ to find the sum over the $t$--dependent terms, giving

\begin{equation}
\left\langle T \right\rangle =-\sum_i \frac{\textrm{Res}\left[ f_\mathcal{P}(z) , z_i \right]}{(1-z_i)^2},
\end{equation}

Similarly, we can find the probability of the process completing \emph{by} $t$, and the variance of the completion times:

%\begin{align}
%p(T\leq t) &= \sum_i\frac{1-z_i^{-t-1}}{1-z_i}\textrm{Res}\left( f_\mathcal{P}(z) , z_i \right), \\
%\textrm{Var}(T) &= \sum_i\frac{1+z_i}{(1-z_i)^3}\textrm{Res}\left( f_\mathcal{P}(z) , z_i \right) +\nonumber \\ & \phantom{=}\sum_{i,j} \frac{\textrm{Res}\left( f_\mathcal{P}(z) , z_i \right)}{(1-z_i)^2}\frac{\textrm{Res}\left( f_\mathcal{P}(z) , z_j \right)}{(1-z_j)^2},
%\end{align}

\begin{align}
p(T\leq t) &= \sum_i\frac{1-z_i^{-t-1}}{1-z_i}\textrm{Res}\left[ f_\mathcal{P}(z) , z_i \right], \\
\textrm{Var}(T) &= \sum_i\frac{1+z_i}{(1-z_i)^3}\textrm{Res}\left[ f_\mathcal{P}(z) , z_i \right] - \left\langle T\right\rangle^2,
\end{align}

We may note now that, if $f_\mathcal{P}(z)$ is built up constructively, as in Eq.~(\ref{eq:graph_rules_2}), each non-terminating node contributes a single factor of $z$ to the PGF. This means that $f_\mathcal{P}(z)^{-1}$ must be of order $n_\mathcal{P}-\left|\mathbf{I}_\mathcal{P}\right|^2$ at most, and so have no more than $n_\mathcal{P}-\left|\mathbf{I}_\mathcal{P}\right|^2$ poles. When calculating $p_t$ by Eq.~(\ref{eq:p_t_general}), the residues of $f_\mathcal{P}(z)/z$ are not $t$-dependent. Therefore, when we vary $t$, we simply need to calculate $z_i^t$ for each $z_i\in \mathbb{P}(f_\mathcal{P})$. Given again some algorithm for calculating exponentials $k^t$ in $\mathcal{O}(f_\textrm{exp}(t))$ operations, we have that calculating $p_t$ now scales asymptotically as 

\begin{equation} \label{eq:order_mag_mat_mult}
\mathcal{O}\!\left(p_t\hspace{1mm} \textrm{by Cauchy}\right) = \mathcal{O}\!\left(\left[n_\mathcal{P}-\left|\mathbf{I}_\mathcal{P}\right|^2\right] f_\textrm{exp}(t)\right),
\end{equation}

\noindent which represents an improvement of a factor of $n_\mathcal{P}$ over the matrix multiplication method.

%XX Mention that we need one terminating node

%XX talk about interesting non-trivial behaviour coming in only in the presence of closed loops

\section{Error distributions} \label{sec:fft}

%XX make the distinction between heralded processes and not.
In any process, there will be events that have some probability to cause an error. For example, if an event represents a state being stored on a quantum memory, then in each time-step there is some non-zero probability that the memory fails and the information stored on it is lost. When carrying out the process, we wish to know $p(k|t)$; the probability that we will pass such an edge $k$ times, given an overall process completion time of $t$. This implicitly assumes that such a process is \emph{heralded}. That is, we always know what stage of $\mathcal{P}$ we are at, and so can count the number of occurrences of an error-carrying event. A non-heralded process would be one where we have a description of $G_\mathcal{P}$, but we do not know how close we are to completion at any time, but instead are simply informed when the process completes. If each occurrence of an error-carrying event has a probability $\epsilon$ to cause an error, then the overall probability that an error will have occurred is given by 

\begin{equation} \label{eq:p_errors_her_and_non}
\begin{split}
p(\textrm{error heralded}) &= 1-(1-\epsilon)^k, \\
p(\textrm{error non-heralded}) &= \sum_{k=0}^\infty p(k|t)\left[1-(1-\epsilon)^k\right]. \\
\end{split}
\end{equation}

In order to include this in our analysis, we must first identify which events (edges in $G_\mathcal{P}$) may cause an error. Then, for each event in question between edges $j$ and $i$, we multiply $\left[M_\mathcal{P}\right]_{i,j}$ by an open complex variable, $w$, which we will call the counting variable.

Now note that the value for $p_t$, calculated either by Eq.~(\ref{eq:pt_mat_mult}) or Eq.~(\ref{eq:p_t_general}), may be seen as a sum of the probabilities of the different sequences of events by which the process may be completed in time $t$. When one term in $M_\mathcal{P}$ is an open variable, $p_t$ will be expressed as a finite polynomial in $w$, which we will denote $p_t(w)$. For such sequence of events that includes $k$ passes of an error-carrying edge and occurs with probability $p_a$, $p_t(w)$ will include a term equal to $p_a w^k$. The full expression for $p_t(w)$ will then be of the form

\begin{equation} \label{eq:errorProbs}
p_t(w) = p_t(1) \hspace{-3mm} \sum_{k=0}^{\mathcal{O}(p_t(w))} \hspace{-2mm} p(k|t) \hspace{1mm} w^k.
\end{equation} 

\noindent where $\mathcal{O}(p_t(w))$ is the order of $p_t(w)$. Thus by finding the coefficients of this polynomial, we can find the error distributions. This polynomial is finite, with all terms involving $w$ to a non-negative power. Therefore it has no poles, so we cannot use the methods of Section \ref{sec:cauchy}. Instead we can extract the coefficients by way of a (fast) Fourier transform, which, unlike the complex analysis method, can be done numerically. To do this, we first should identify some number $N_\mathcal{P}(t)$ such that $\mathcal{O}(p_t(w)) \leq N_\mathcal{P}(t) \leq 2 \mathcal{O}(p_t(w))$, where the latter inequality is to avoid aliasing effects \footnote{It may seem like we can always choose $N_\mathcal{P}(t)=t$. However, if $M_\mathcal{P}$ is constructed from elementary process matrices by Eq.~(\ref{eq:mP_tensor_or}) or Eq.~(\ref{eq:mP_tensor_and}) we may need to choose a larger value for $N_\mathcal{P}(t)$}. We then evaluate $p_t(w)$ at $N_\mathcal{P}(t)$ equally spaced complex points, given by $\big\{ p_t(e^{i 2 \pi k/N_\mathcal{P}(t)}) \big| k=1,\dots,N_\mathcal{P}(t) \big\}$. The discrete Fourier transform of these evaluated points reveals the first $N_\mathcal{P}(t)$ coefficients of $p_t(w)$ (where all greater coefficients are 0). Applications of this Fourier method for extracting coefficients to more general analytic functions are described in \cite{fornberg1981numerical}.

This construction may also be used to account for different kinds of errors, by multiplying matrix elements by different complex variables, $w_1, w_2, w_3, \cdots$, and performing a multi-dimensional Fourier transform on $p_t(w_1,w_2,w_3,\cdots)$ to determine $p(k_1,k_2,k_3,\cdots|t)$.

From this we can also read off the \emph{average} error rate for a process completing by time $t$. i.e. the non-heralded error. Suppose we have only a single type of error. From Eqs.~\ref{eq:p_errors_her_and_non} and \ref{eq:errorProbs}, we can say that

\begin{equation}
p(\textrm{error non-heralded}) = \frac{1-p_t(w=1-\epsilon)}{p_t(1)}
\end{equation}

The case for multiple types of error follows as a simple extension of this.

\section{Innsbruck protocol analysis} \label{sec:innsbruck}

\begin{figure}[t]
   \centering
     \includegraphics[width=.8\columnwidth]{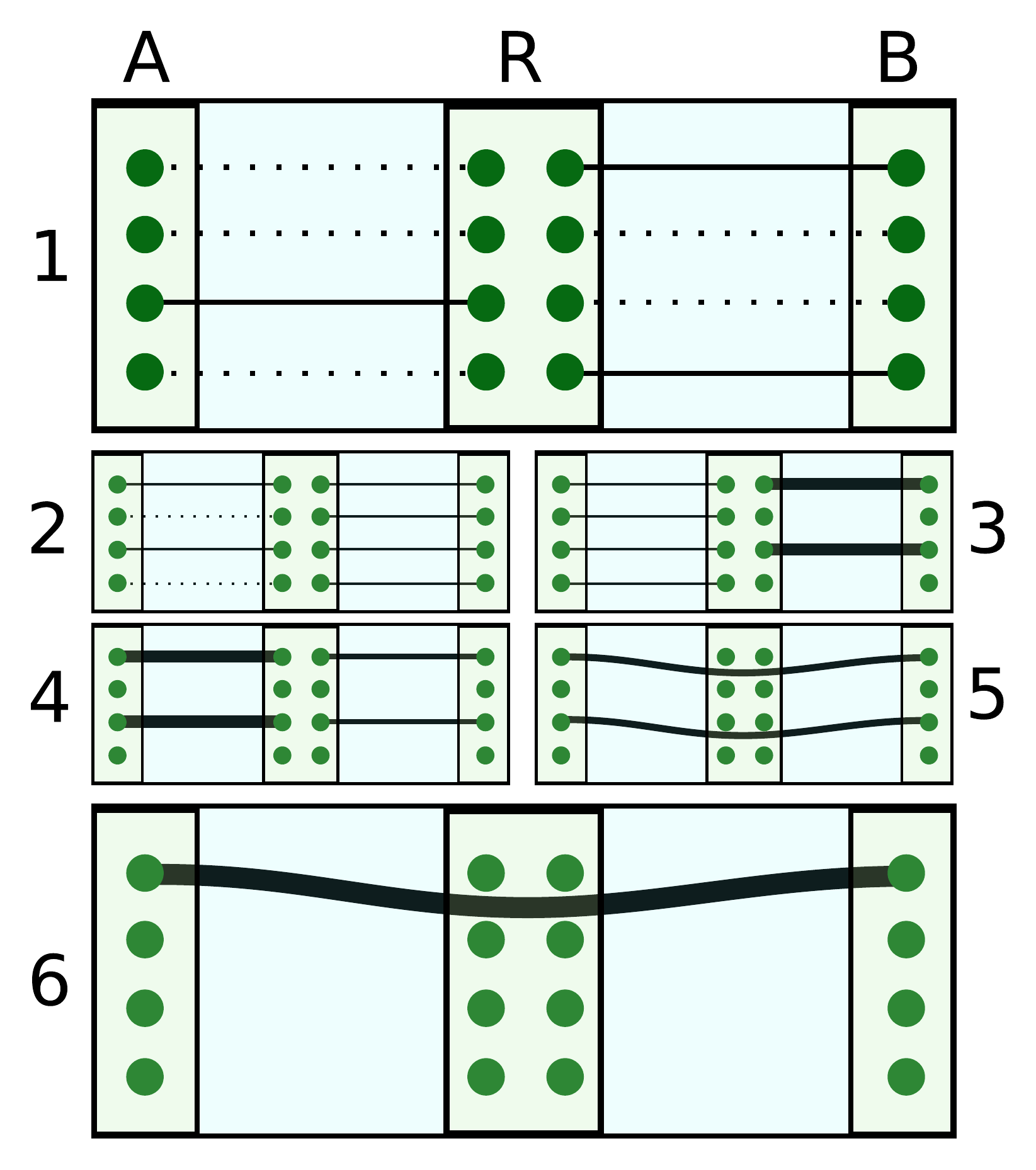}
     \captionsetup{justification=justified}
     \caption{Innsbruck quantum repeater scheme with 2 sections. This shows the entanglement generation scheme for one ``bunch,'' (defined in the last paragraph of Section \ref{sec:intro}) where $q_0=4$. Part 1 shows a situation where some pairs of memories have established entanglement, where $A$ is Alice, $R$ is Richard and $B$ is Bob. Circles show quantum memories dotted lines show no entanglement and solid lines are successful entanglement connections. In parts 2--3, the right-hand section has completed, and distilled down to $q_1=2$ pairs. The thickness of the lines between memories shows the fidelity of that entangled state. In parts 4--5, the left-hand section is also distilled and these states are connected to the right-hand pairs by entanglement swapping. In part 6 the two pairs between Alice and Bob are distilled again to form a single high-fidelity pair.}
     \label{fig:inns_schem}
\end{figure}

In this section we will be considering the repeater protocol of the Innsbruck group \cite{Briegel1998}. In the standard implementation of this protocol, there are $Q_0$ pairs of quantum memories between each adjacent pair of repeater stations. These are all connected in parallel, and then distilled to make $Q_1\leq Q_2/2$ pairs. By entanglement swapping, these are then connected with adjacent pairs to form entanglement over twice the length, and distilled again to form $Q_2\leq Q_0/4$ pairs in parallel between each section, and so on. Previous analyses of this protocol have either assumed that the entanglement connection can be done almost deterministically, or considered that entanglements establish after some average time. So if each attempt to establish entanglement between two stations has a probability to succeed of $p$, then a simplified approach to understanding the system and estimating the key rate would be to assume that all pairs establish entanglement after $1/p$ attempts. 

Considerable progress has been made in understanding and building upon the Innsbruck protocol, since it is one of the most promising routes to constructing long-distance quantum communication. Much of this work has focused on aspects such as the relation between the key rate and experimental imperfections \cite{abruzzo2013quantum}, the specifics of how to implement the system with atomic ensembles \cite{Duan2001}, understanding and improving the robustness against channel noise \cite{Sangouard2011QuantumOptics,Vinay2017PracticalCommunication} or side-channel attacks \cite{lydersen2010hacking,lamas2007breaking,nauerth2009information,Vinay2018}. However, one important aspect is often overlooked, namely the statistical factor of waiting times arising from probabilistic completion times of different elements. We now show that this has severe implications for the performance of the protocol.

In our analysis, we let the $Q_0$ parallel pairs of memories between each pair of repeater stations be divided into bunches of $q_0$ pairs. When all pairs within such a bunch have completed, then they are distilled to $q_1 \leq q_0/2$ pairs. This is an inequality, since distillation (as described more fully in Appendix \ref{ap:process}) is a probabilistic process. This is shown in Fig.~\ref{fig:inns_schem}. Here, there is a trade-off inherent in the size of $q_0$. When $q_0$ is small, the bunch will complete quickly on average. This means that the first entangled pair to complete will not have to wait long before the last one completes, and so is less likely to accrue memory errors. However, one then has fewer options for distilling a high fidelity state. Given a large set of states, we can instead find a better optimal strategy for combining states under a distillation protocol to result in a higher final secret key rate, at the expense of longer waiting times.

\subsection{Constructing the matrix} \label{sec:constructing_inns_mat}

Here we consider at first a repeater consisting of two sections, separated by a distance $L$. Alice tried to establish entanglement between herself and Richard (a repeater station), and Richard between himself and Bob. The Markov graph for the establishment of a single Bell pair is shown in Fig.~\ref{fig:GP}. Let its Markov matrix be $M_\textrm{Bell}$. We shall consider one time-step in this process to be $2L/c$, where the factor of 2 arises since the receiving party needs to send a classical signal back to the sending party to confirm whether the previous photon was received or not.

We now use Eq.~(\ref{eq:mP_tensor_or}) to construct the matrix for $q_0$ pairs connecting in parallel between a pair of repeater stations. We want to include a complex counting variable, $w_0$ that counts how many time-steps a given quantum memory has to wait before the others finish. However, we should note that we include this only on a single factor of the matrix for the section, $M_\textrm{sect}$, to avoid multi-counting errors. The counter $w_0$ therefore counts how many errors accumulate on a particular entanglement link. By symmetry, we can say that this error distribution is equal across all such entanglement links. Therefore,

\begin{equation} \label{eq:M_sec}
\begin{split}
M_\textrm{sect} = &\left[M_\textrm{Bell}+w_0 \hspace{1mm}\textrm{diag}(\mathbf{I}_\textrm{Bell}\right] \otimes\\& \left[M_\textrm{Bell}+\textrm{diag}(\mathbf{I}_\textrm{Bell})\right]^{\otimes q-1} \\&- w_0\hspace{1mm}\textrm{diag}(\mathbf{I}_\textrm{Bell}^{\otimes q}),
\end{split}
\end{equation}

\noindent where

\begin{equation}
M_\textrm{Bell} = 
\begin{bmatrix}
    1-p  & 0\\
    p    & 0 
\end{bmatrix}, \hspace{2mm}
\mathbf{I}_\textrm{Bell} = 
\begin{bmatrix}
0\\
1
\end{bmatrix}.
\end{equation}

\noindent where $p$ is the probability of entanglement being established in any particular attempt. In Appendix \ref{ap:simple} we describe how this may be reduced in dimension by symmetry arguments. 

Once all pairs between two stations have established their entanglement, we want to perform a distillation on these. These are matched up into $\left\lfloor{q_0/2}\right\rfloor$ pairs, which are then distilled using the DEJMPS protocol \cite{Deutsch1996QuantumChannels}. A DEJMPS distillation between two noisy Bell pairs has some non-unity chance of success, which depends on the fidelities of the states involved. However if the success probabilities were fidelity-dependent, that would mean including terms in the matrix which depend on the time taken for the process to reach that event, a modification which would move us outside the realm of Markovian dynamics. Therefore we will choose some minimum distillation success probability, $\lambda$, corresponding to the success probability two states of fidelity $F_\textrm{min}$ being distilled with each other. We will later exclude any runs of the process that would have used states of fidelity less than $F_\textrm{min}$, as explained in Appendix \ref{ap:process}. In this way, any choice of $\lambda$ will give us a lower bound on the secret key rate, and we may freely maximise over choices of $\lambda$. We therefore add a row and column to $M_\textrm{sect}$ to form $M_\textrm{dist}$. The new column has an element representing distillation success, with a probability of $1-(1-\lambda)^{\left\lfloor{q_0/2}\right\rfloor}$. The ``failure'' event (of all distillations failing) resets the process of creating entanglement on that section. Therefore, if $M_\textrm{sect}$ in Eq.~(\ref{eq:M_sec}) takes the form

\begin{equation} \label{eq:M_sec_general_block}
M_\textrm{sect} = 
\begin{bmatrix}
M_\textrm{sect}^\prime & \mathbf{0} \\
\mathbf{a}^T & 0
\end{bmatrix},
\end{equation}

\noindent for some matrix $M_\textrm{sect}^\prime$ and vector $\mathbf{a}$, then $M_\textrm{dist}$ is given by

\begin{equation} \label{eq:M_dist}
M_\textrm{dist} = 
\begin{bmatrix}
M_\textrm{sect}^\prime & \mathbf{b} & \mathbf{0} \\
\mathbf{a}^T & 0 & 0 \\
\mathbf{0}^T & s & 0 
\end{bmatrix},
\end{equation}

\noindent where the vector $\mathbf{b} = \left[ (1-\lambda)^{\left\lfloor{q_0/2}\right\rfloor}, 0, 0, \cdots, 0 \right]^T$, and ${s=1-(1-\lambda)^{\left\lfloor{q_0/2}\right\rfloor}}$. 

Finally, we construct the matrix for the entire system, $M_\textrm{full}$, in a similar way to Eq.~(\ref{eq:M_sec}) by considering two copies $M_\textrm{dist}$. We again include a complex counting variable to account for all memories on one section needing to wait until the other side has been connected and distilled. For this we make sure to use a different counting variable, $w_1$, so we can keep track of the distribution of errors that occur before and after the first round of distillation.

\subsection{Analysis}

We can now analyse $\{p(k_0,k_1|t)\}$, where $k_0$ and $k_1$ are the number of passes of edges weighted by $w_0$ and $w_1$ respectively. By doing this for a fixed $q_0$, we can find a distillation strategy that gives the maximum possible achievable secret key rate for a given completion time, $t$, averaged over the error distribution (explained in detail in Appendix \ref{ap:process}), which we shall call $K(t|q_0,p,\epsilon_W)$. Since the key rate of a protocol goes inversely with the time taken to establish a raw bit, and linearly with the number of parallel ``bunches'' of states that are used, $Q_0/q_0$, we will use the normalised average key rate as a function of $q_0$ as our main figure of merit for the system analysis:

\begin{equation}\label{eq:K(q)}
K(q_0,p,\epsilon_W) = \frac{1}{q_0}\sum_{t=1}^\infty \frac{p_t}{t} K(t|q_0,p,\epsilon_W)
\end{equation}

%XX note discuss timings, one is longer than others but doesn't make a difference.
%XX \epsilon_W = 1-exp(4L/cT), T = half-life of quantum memory, extra factor of 4 comes from 2 memories involved.
%XX Exponential growth of matrix size with number of Sections.
%XX discard a greater proportion of states
%XX if q is odd, we leave on out of the first distillation

\begin{figure}[t]
   \centering
     \includegraphics[width=\columnwidth]{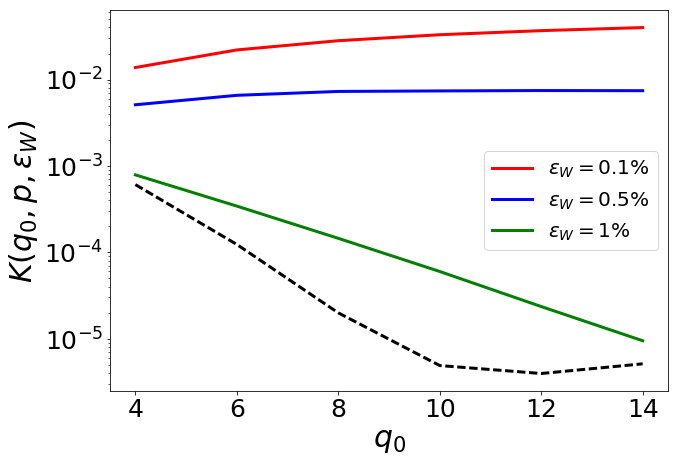}
     \captionsetup{justification=justified}
     \caption{The normalised secret key rate, in the case where $p=0.1$. Plots are in the same order as in the legend. Solid lines show the key rate including statistical effects ($K$) and the dashed line shows the simplified key rate ($K_\textrm{simp}$), which is identical for all $\epsilon_W$.} %XX remove boundary of legend. 
     \label{fig:skr_inns}
\end{figure}

For the error probability per time period, $\epsilon_W$, we assume the probability for a quantum memory to not undergo an error decays exponentially with time as ${\epsilon_W = 1-\exp(4L/c\tau)}$, where the extra factor of 2 is due to the fact that each entangled pair involves 2 memories. Here, $\tau$ is the memory lifetime. In Fig.~\ref{fig:skr_inns} we show the a few examples of calculated normalised key rates for different values of $q_0$. In order to compare the insight gained from this method to the estimations that might result from a less nuanced analysis, we have also shown the \emph{simplified secret key rate}. Here, we consider only the average connection time of an entanglement link. That is to say, we assume that all links wait for a time $1/p$, and then connect deterministically. For a fair comparison, we have retained the same maximization over distillation strategies that is outlines in Appendix \ref{ap:process}. The simplified secret key rate may therefore be written as

\begin{equation}
K_\textrm{simp}(q_0,p,\epsilon_W) = \frac{1}{q_0}\sum_{t=1}^\infty \frac{p_t}{t+1/p} K(t|q_0,1,\epsilon_W)
\end{equation}

When we consider Fig. \ref{fig:skr_inns}, we can see that an estimation of the secret key rate that considers only the average completion time severely underestimates the performance of the protocol. This is particularly striking when we note that no errors accumulate in the simplified analysis, due to the fact that no elements are left waiting while others complete. In particular, the difference between the statistical and simple key rates in the $\epsilon_W = 1\%$ case reaches 3.8 orders of magnitude, which could mean the difference between communicating in kilobits and megabits per second.

We may also note that, for low values of $\epsilon_W$, the key rate rises with increased bunch size. This means that increasing the pool of states available to be distilled has a greater-than-linear effect on the key rate, highlighting the power and importance of distillation to quantum technologies.

\section{Analytic Key Rate Bounds for the Innsbruck protocol} \label{sec:akrb}

The techniques presented above allow for a thorough investigation of the contribution of statistical factors arising from non-deterministic protocol elements towards the secret key rate of a general quantum communication protocol. While this has been presented in-depth for a two-section repeater, practical systems will often demand the application of a series of many repeaters. The current limit for repeater-unassisted quantum communication is on the order of a hundred kilometers. If we therefore wish to securely communicate on an intercontinental scale, we require a method of analysis that can scale up to dozens of repeater sections. This presents a limitation in our protocol: while the dependence of $n_\mathcal{P}$ on $q_0$ can be made linear (Appendix \ref{ap:simple}), the dependence on the number of sections, $N_S$, remains exponential.

In the original analysis of the Innsbruck protocol, the fidelity of the final shared state was not considered to be fundamentally dependent on the number of sections. This is because the $l^\textrm{th}$ level of the protocol, which consists of taking entangled pairs over some distance $2^lL$, distilling them, and connecting with adjacent sections to form pairs over a distance $2^{l+1}L$, would produce pairs of a fidelity that did not depend on $l$. 

In addition to waiting times increasing with $q_0$, there is also the issue that the classical communication time grows with the distance over which pairs are entangled. However, if statistical factors are ignored then this could be dealt with using a ``blind'' protocol, where distillation and attempts are assumed to have succeeded at every stage, and communication after the termination of the protocol allows for a post-selection on the attempts that succeeded. This allows for the final fidelity to be kept above the minimum level required for secure communication at the expense of a hit to the raw key bit generation rate. %XX Is above true?

Once we include the statistical waiting times in our analysis, it is no longer true that a non-zero secret key rate can be guaranteed for all length scales. In this section we examine the behaviour of a repeater network with a minimum requirement of physical qubits, which is $q_0=2^{N_S}$. Instead of finding best key rates, we look here for the parameters for which the secret key rate is lower bounded above zero. By finding the threshold parameters that need to be reached for the protocol to operate, we can identify concrete values for component designers to aim for, and give benchmarks by which we can compare performances.

We will consider that after every quantum operation, a Werner twirling procedure \cite{bennett1996purification} is applied to all states. This involves unitarily mapping the state to the singlet state, $\ket{Psi^-}$, then applying a randomly chosen local unitary Pauli operation identically each part of the entangled pair. This maps all states to Werner states of the same fidelity (see Appendix \ref{ap:process}). By doing this, we can simply consider the effect of the repeater network as a recursive function on a single real variable - that of the average fidelity. Note that this operation is not actually carried out, it is simply used to repeatedly map states to the analytically simple Werner states. This may be done since applying local operations cannot increase the strength of entanglement by any measure, and so cannot increase the secret key rate. %XX cite?

The analysis then proceeds as follows. Level 0 of the protocol consists of all pairs within one section connecting at initial fidelity $F_\textrm{init}$. Instead of considering the full probability distribution of waiting times, we consider that all pairs wait for a number of time-steps equal to the estimated time for the last pair to connect, $k_{L}$, which upper bounds the waiting time for each pair. This time is equal to the expectation value for the largest order statistic from a sample of $q_0$ chosen from the distribution with cumulative distribution function $1-(1-p)^t$. By choosing $p\ll 1$, such that we may allow the distributions to be approximated by continuous functions, we may use results from \cite{arnold1979bounds} to bound this by

\begin{equation}
k_{L} \leq \left(\frac{q-1}{\sqrt{2q-1}}+1\right)\frac{1}{\left|\log(1-p)\right|}+1.
\end{equation}

These states are then distilled to produce states of fidelity

\begin{equation}
F_0 = J[D(F_\textrm{init},\epsilon_W,k_{L})],
\end{equation}

\noindent where the functions describing the effect of the decay of quantum memories over time $k_{L}$ on the average fidelity and DEJMPS distillation are given respectively by

\begin{equation}
\begin{split}
D(F,\epsilon_W,{k_{L}}) &= (1-\epsilon_W)^{k_{L}}F + \frac{1-(1-\epsilon_W)^{k_{L}}}{4},\\
J(F) &= \frac{10F^2 - 2F + 1}{8F^2 - 4F + 5},
\end{split}
\end{equation}

\noindent respectively. The $l^\textrm{th}$ level of the protocol consists of the following when $l \geq 1$. Within each pair of two sections, one section will complete first, and wait for a time no longer than $k_{A,l}$ for the latter to complete. We show in Appendix \ref{ap:tau} that this is bounded by

\begin{equation} \label{eq:harmonic_eq}
k_{A,l} \leq 2^l\left[\frac{H\!\left(2^{N_S-l+1}\right)}{\left|\log(1-p)\right|}+1\right],
\end{equation} %2^{N_S-l} = 2q_l where q_l=2^{N_S-l}

\noindent where $H\!\left[n\right] = \sum_{m=1}^n 1/m$ is the $n^\textrm{th}$ harmonic number. The average fidelity after level $l$ can then be defined recursively as

\begin{equation}
F_l = J(C(F_{l-1}, \tilde{F}_{l-1}, \epsilon_L)),
\end{equation}

\noindent where $C$ gives the average fidelity after connecting two adjacent sections and twirling, where we have allowed here for local gate errors. This is given by 

\begin{equation}
C(F_a, F_b, \epsilon_L) = D\left(\frac{1}{3}(1-F_a)(1-F_b)+F_a F_b, \epsilon_L, 1\right),
\end{equation}

\noindent where $\epsilon_L$ is the probability that an error occurs when performing the local operations involved in entanglement swapping, and $\tilde{F} = D(F_{l-1}, \epsilon_W, k_{A,l-1})$.

\begin{figure}[t]
   \centering
     \includegraphics[width=\columnwidth]{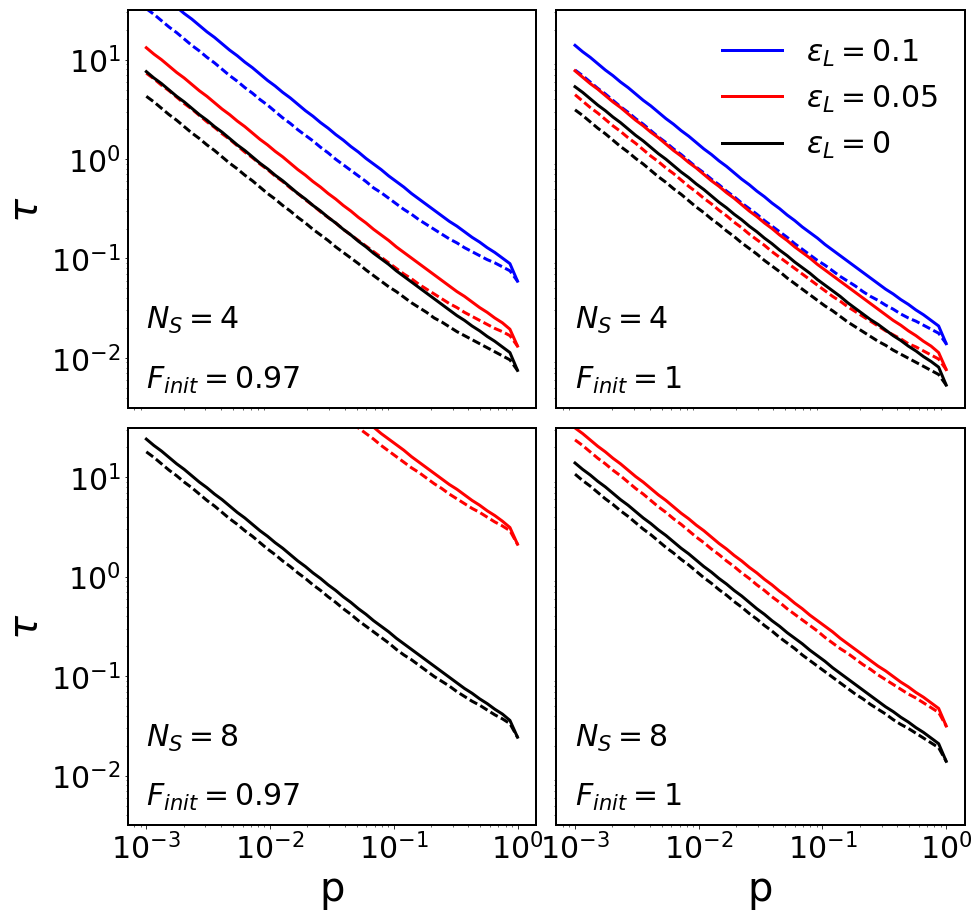}
     \captionsetup{justification=justified}
     \caption{Minimum quantum memory lifetime, $\tau=-4L/\log(1-\epsilon_W)c$, required for a minimum-resources quantum repeater to securely communicate over a network of 8 sections of length $L=25\textrm{km}$. $F_\textrm{init}$ is the initial fidelity of entangled pairs before distillation or connection, and $N_S$ is the number of sections. Solid lines indicate the cases that include statistical factors, and dashed lines do not. Both solid and dashed lines are ordered the same in the plots and the legend. Blind distillation is not used in either case.}
     \label{fig:analyticInns}
\end{figure}

In Fig.~\ref{fig:analyticInns} we show the minimum quantum memory lifetimes required for a non-zero key rate as a function of $p$ From this it can be seen that the probability for an entanglement attempt to succeed is the biggest factor in affecting the ability to securely communicate. For comparison we also include the requirements for the case that does not include statistical effects, where $k_{A,l}=2^l$ for all $l$. It can be seen that the minimum memory requirements are slightly higher in the case where statistical effects are included, but this effect decreases with the number of sections over which we connect. For an even comparison, we have not used blind distillation in the non-statistical case. We see that there is a constant-factor increase in the required lifetime of the memories. In some cases this reaches as high as a factor 2 increase in the required lifetime of the quantum memories. The resultant bounds are just reachable by the lifetimes of atomic ensembles, which can have lifetimes up to 40ms \cite{krauter2011entanglement}. However, all bounds are well within the lifetimes of the nuclear spin states of NV centers \cite{Ladd2005CoherenceSilicon}. This implies that the main challenge towards implementation of DLCZ-type protocols \cite{Duan2001Long-distanceOptics.} is the construction of optical elements with high transmission and detection efficiencies, whereas NV-center-based protocols \cite{Nemoto2014PhotonicCenters,Vinay2017PracticalCommunication} may be more suitable when these efficiencies are low.

%XX For a bound, only need to know error rate - assume all connect
%XX Longer sections have to wait for a longer time, multiply waiting time by (2^l).
%XX Convert to Werner state - Werner twirling
%XX specify this is a just-enough repeater - minimum physical reqs.
%XX replace taus here with ks for time step numbers, and be clear that they are time steps

\section{Conclusion}
  Many of the practical quantum technologies that are being proposed are inherently probabilistic in nature, which leads to uncertain completion times and error distributions. We have developed techniques that allow for thorough characterizations of such statistical distributions in both the computational and analytic directions. In terms of computational techniques, we have used Markov chains to analyse quantum entanglement generation. We have shown how to form composite systems from smaller elements in a constructive manner. We have then shown how to use the Markov matrices for such composite systems to calculate the probability distribution over the number of errors that occur in the running of a general protocol. This allows for a complete characterization of the fidelities of the states that are produced by a quantum protocol. As an example, we have analysed the Innsbruck quantum repeater protocol with a memory-error model. A thorough understanding of the set of resultant errors has been shown to lead to a tighter bound on the secret key rate than an analysis based only on an averaged approach. In some cases this resulted in tightening the bounds on key rates by over 3 orders of magnitude -- a clear indication that a consideration of statistical effects does not simply provide a minor correction to performance, but instead is fundamental to understanding the quantitative behaviour of a system.

In terms of analytic techniques, we have shown how elements of the eigenvectors of a transformed form of the Markov matrix correspond to the probability generating function of the process. This has been solved for an arbitrary term in the probability distribution over completion times by using results from complex residue analysis. This was done in a way such that the number of computational operations required scales only linearly with the number of states in the Markov process, compared to the quadratic scaling of a more direct approach. Finally we have shown how the theory of order statistics can put bounds on the statistical effects on the secret key rate, and used this to bound the minimum quantum memory lifetimes needed to run the Innsbruck repeater protocol.

\section*{Acknowledgements}

This research was made possible via the EPSRC Quantum Communications Hub, Grant No. EP/M013472/1

\appendix

\section{Simplifying Markov matrices with high symmetry} \label{ap:simple}
\begin{figure}[t]
   \centering
     \includegraphics[width=\columnwidth]{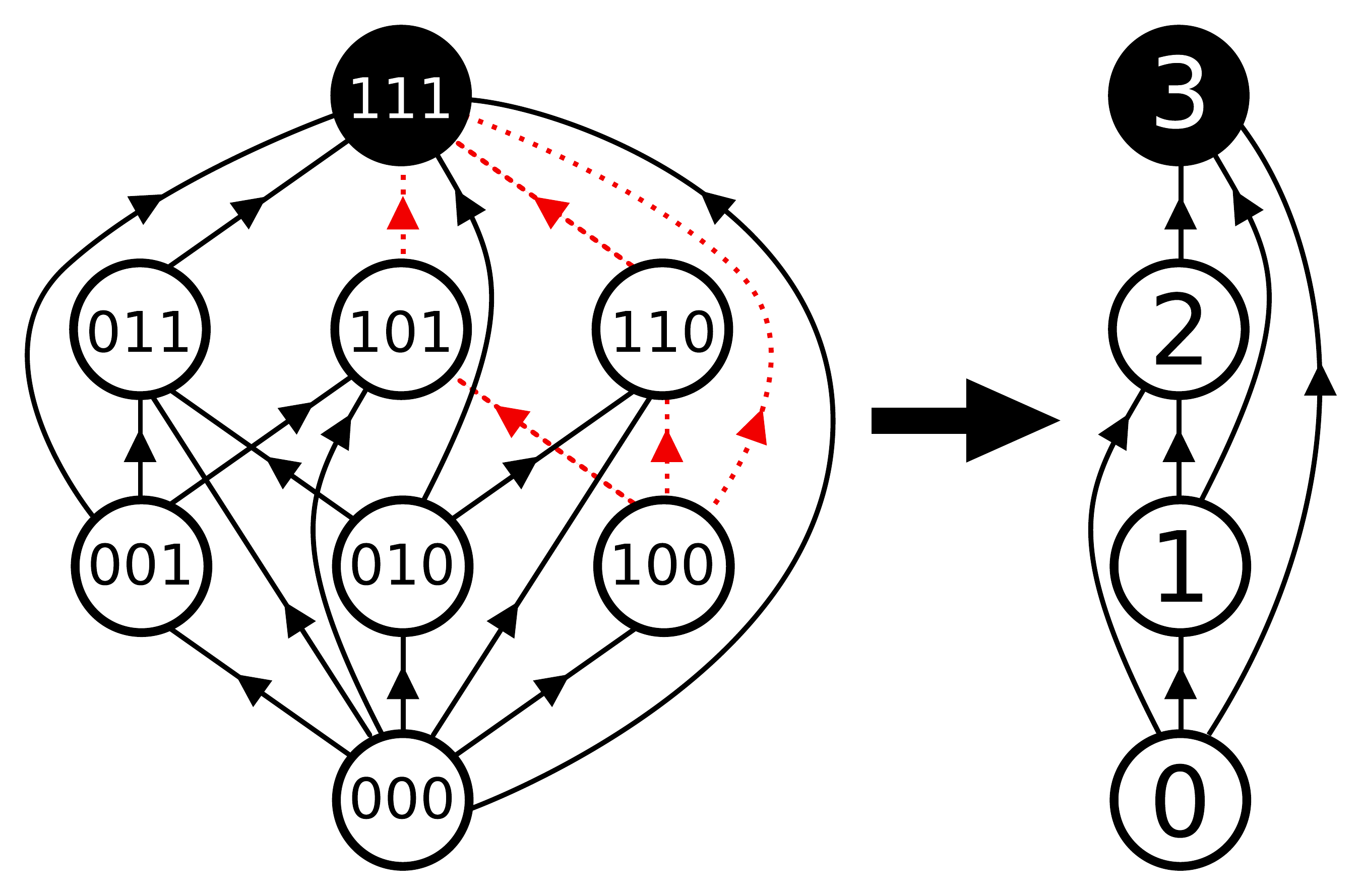}
     \captionsetup{justification=justified}
     \caption{On the left we have the Markov matrix for 3 pairs of entanglement trying to establish in parallel with transition probabilities not shown for clarity, constructed in a way that tracks the status of each pair. The binary codes on each state show whether the first, second and third pairs are connected (1) or unconnected (0). Shown as red dotted lines are the events involve the first pair waiting after its completion. These edges translate to terms in the matrix that should be multiplied by an error-counting variable, $w$. On the right, we have grouped the states by how many pairs are connected. Terminating nodes shown in black. If the probability for each unentangled pair to establish its entanglement in a given time-step is $p$, then the probability to transition from node $j$ to node $i$ in the lumped process after application of the mixing matrix is given by ${{q-j}\choose{q-i}} p^{i-j} (1-p)^{q-i} [q+(w-1)j]/q$.}
     \label{fig:lumping}
\end{figure}

The Markov matrix for a single section given in Eq.~(\ref{eq:M_sec}) is {$2^{q_0}$-dimensional}. While this accurately describes the dynamics of the system, we can take advantage of the fact that the system contains a high degree of symmetry to reduce the size of the matrix. We can use the fact that the probability to move between one state and another is only dependent on \emph{how many} entanglements have been established in the initial and final states, and not on the specifics of \emph{which} entanglements. We can therefore use a technique called ``lumping,'' where we create a partition of the states into sets, as shown in Fig.~\ref{fig:lumping} (discussed in more detail in \cite{kemeny1983finite}). From this we can consider a new process, where each set of states is considered as a single state.

When we lump states together, we should ensure that the transition probabilities in the lumped process produce the same system behaviour as in the unlumped process. Let $M$ be the (unlumped) Markov matrix for the process, and $A_1, A_2, \cdots$ be a partitioning of the states, where each $A$ is a set of states disjoint from all other sets. Then in order to be able to lump the process we require that, for each $A_m$ and $A_n$, $\sum_{i\in A_m} \!M_{i,j}$ should be identical for all $j\in A_n$.

For our system this is true when we simply consider the transition probabilities, but the symmetry is broken when we include the complex counting variable, $w$, since this is only applied to one of the pairs. However, we can re-introduce a symmetry here, since $w$ is designed to capture the error rate on a typical pair, and not a particular pair. We pre-multiply the original Markov matrix by an \emph{in-set maximal mixing matrix}, $M_\textrm{mix}$, which takes us from some state to any other state with the same number of completed entangled pairs with equal probability. This is given by a block diagonal, where each block has all elements equal to $1/n$, where here $n$ is the size of the block. This is shown for the $q_0=3$ case below.

\begin{equation}
M_\textrm{mix} = \frac{1}{3}
\begin{bmatrix}
3 & 0 & 0 & 0 & 0 & 0 & 0 & 0 \\
0 & 1 & 1 & 1 & 0 & 0 & 0 & 0 \\
0 & 1 & 1 & 1 & 0 & 0 & 0 & 0 \\
0 & 1 & 1 & 1 & 0 & 0 & 0 & 0 \\

0 & 0 & 0 & 0 & 1 & 1 & 1 & 0 \\
0 & 0 & 0 & 0 & 1 & 1 & 1 & 0 \\
0 & 0 & 0 & 0 & 1 & 1 & 1 & 0 \\
0 & 0 & 0 & 0 & 0 & 0 & 0 & 3
\end{bmatrix}
\end{equation}

This effectively distributes the counting variable amongst the states, giving it the symmetry required to lump the states together. This reduces the number of unique states in the process corresponding to one section from $2^q$ to $q+1$.

\section{Detailed description for the analysis of the Innsbruck protocol} \label{ap:process}

The first step in analysing the modified Innsbruck protocol is to construct the associated Markov matrix, as described in Section \ref{sec:constructing_inns_mat} and Appendix \ref{ap:simple}. To do this we fix $q_0$, which sets the size of the matrix, and $p$, which determines the elements of the matrix. For a fixed $t$, a joint distribution of $k_0$ and $k_1$ is then calculated. 

We assume that the states that are initially created after establishment is connected are Werner states of the form

\begin{equation} \label{eq:WernerPsi}
\rho = \frac{4F_\textrm{init}-1}{3}\ket{\Phi^+}\bra{\Phi^+}+\frac{1-F_\textrm{init}}{3}\mathbb{1}.
\end{equation}

\noindent where $F_\textrm{init}$ is the fidelity with respect to $\ket{\Phi^+}$. We then choose some $k_\textrm{max}$ that gives some largest acceptable error. Then from the marginal distribution on $k_0$ we then choose $2q-1$ values for $k_0$ (all of which are below $k_\textrm{max}$), and we choose a final value for $k_0$ and a value for $k_1$ from the full error distribution. These transform the $2q$ states (by $q$ on each section of the repeater network) as:

\begin{equation}\label{eq:error_evolution}
\rho \mapsto (1-\epsilon_{W0})\rho + \epsilon_{W0}\mathbb{1},
\end{equation}

\noindent where $\epsilon_{W0}$ are defined as heralded errors as in Eq.~(\ref{eq:p_errors_her_and_non}), with $k=k_0$.

We now partition the set of states into a `left set' and a `right set', corresponding to the two different section of the network, and randomly apply a distillation to each set. To do this, we pair up the states within a set. If $q_0$ is odd, one state is randomly chosen to proceed to the next round without being distilled. The remaining pairs are distilled according to the DEJMPS protocol, which maps states on the basis $\ket{\Phi^+}, \ket{\Psi-}, \ket{\Psi^+}, \ket{\Phi^-}$, with diagonal coefficients $(a_1, b_1, c_1, d_1), (a_2, b_2, c_2, d_2)$ to the state with coefficients 

\begin{equation}
\frac{1}{\mathcal{N}}
\begin{bmatrix}
a_1a_2+b_1b_2\\
c_2d_1+c_1d_2\\
c_1c_2+d_1d_2\\
a_1b_2+a_2b_1
\end{bmatrix},
\end{equation}

\noindent where the distillation succeeds with probability $\mathcal{N}=(a_1+b_1)(a_2+b_2)+(c_1+d_1)(c_2+d_2)$. %XX Clarify we do not actually twirl, so have Phi+ again

As explained in the main text, when translating the distillation success probability to a term in the Markov matrix, we use a constant probability of distillation success, $\lambda$. This is related to our choice of $k_\textrm{max}$ by

\begin{equation}
k_\textrm{max} = \left\lfloor\log\left(\frac{3\sqrt{2\lambda-1}}{4F_\textrm{init}-1}\right)\frac{1}{\log{(1-\epsilon_{W0})}}\right\rfloor
\end{equation}

\noindent Two Werner states that have waited for $k_\textrm{max}$ will be of fidelity $F_\textrm{min}$. If these are distilled with each other the success probability will be no less than $\lambda$.

After the states on each section, the number of remaining states on each side, $q_1^L$ and $q_1^R$, are random variables, with $p(q_1^{L,R}=x) = \lambda^{x}(1-\lambda)^{q_0/2-x}$. When we perform entanglement swapping to connect the two sections, the final number of states will be $q_1=\textrm{min}(q_1^L,q_1^R)$ with 

\begin{equation}
\begin{split}
p(q_1=x) =\hspace{1mm}& p(q_1^L=x)\cdot\sum_{y=x}^{q_0/2}p(q_1^R=y) \hspace{1mm}+\\
&p(q_1^R=x)\cdot\sum_{y=x}^{q_0/2}p(q_1^L=y) \hspace{1mm}-\\& p(q_1^L=x)\cdot(q_1^R=x)
\end{split}
\end{equation}

One of the two sets only then undergoes waiting errors while waiting for the other side to complete, by evolving according to Eq.~(\ref{eq:error_evolution}) but with the $\epsilon_{W1}$ calculated from $k_1$.

For a fixed $q_1$, we then calculate the secret key rate as follows. We choose a random pairing of states on the left with states on the right. They are deterministically connected by applying a CNOT gate to the part of each Bell state stored in the repeater, and then measuring each in the $X$ basis. This maps two states of diagonal coefficients $(a_1, b_1, c_1, d_1), (a_2, b_2, c_2, d_2)$ to one with coefficients

\begin{equation}
\begin{bmatrix}
a_1a_2 + b_1b_2 + c_1c_2 + d_1d_2\\
a_1b_2 + a_2b_1 + c_1d_2 + c_2d_1\\
a_1c_2 + a_2c_1 + b_1d_2 + b_2d_1\\
a_1d_2 + a_2d_1 + b_1c_2 + b_2c_1
\end{bmatrix}.
\end{equation}

These final states may then be distilled again. We optimize over combinations of distillation pairings to produce $q_2\leq q_1$ final pairs, in order to maximise the secret key rate, given by

\begin{equation}
K(t|q_0,q_1,p,\epsilon_W) = \sum_{i=1}^{q_2}1-2h_2(\tilde{\epsilon}_i),
\end{equation}

\noindent where $h_2(x) = -x\log_2(x)-(1-x)\log_2(1-x)$ is the binary entropy function, and $\tilde{\epsilon}_i$ is the bit error of the $i^\textrm{th}$ entangled pair, averaged between measuring in the $Z$ basis and the $X$ basis.

We must finally multiply $p_t$ by the probability that none of the states involved in completing the process were of a fidelity less than $F_\textrm{min}$. As such, we make the transformation

\begin{equation}
p_t \mapsto p_t \times \left(\frac{1}{p_t}\sum_{k=k_\textrm{max}}^t \!p(k|t)\right)^{q_0}.
\end{equation}

This key rate is optimized over distillation strategies (both before and after entanglement-swapping) and entanglement-swapping pairing choices, and averaged over values of $q_1$ and selections of sets of $k_0, k_1$ from the distribution to get $K(t|q_0,p,\epsilon_W)$, which is used to get $K(q_0,p,\epsilon_W)$ by Eq.~(\ref{eq:K(q)}).

\section{Analytic calculation of average waiting times} \label{ap:tau}

%XX Update to include extra classical comm waiting time, expand

The distance over which communication has to occur at level $l$ scales with $2^l$. Given two sections of a repeater, there is some number of time-steps $k_\textrm{2sec}$ between the first completing and the second. After the second section completes, there must be one more round of classical communication to indicate this fact. Therefore $k_{A,l} = 2^l(k_\textrm{2sec}+1)$.

We wish to calculate $k_\textrm{2sec}$, which is given by $\mathbb{E}[\left|x-y\right|]$, where $x$ and $y$ are two times drawn from the distribution, $f(t)=\textrm{d}_t\hspace{.5mm}C(t)$, where the cumulative distribution function is given by $C(t)=[1-(1-p)^t]^q$. Approximating these as continuous distributions, we can write this as

\begin{equation}
\begin{split}
\mathbb{E}[x-y|x>y]\hspace{1mm}+\hspace{1mm}&\mathbb{E}[y-x|y>x] =\\
&2\int_0^\infty \int_0^y (y-x)\hspace{1mm} f(x)\hspace{1mm} f(y)\hspace{1mm} \textrm{d}x \hspace{1mm}\textrm{d}y.
\end{split}
\end{equation}

\noindent Let this inner integral be $I$. Then 

\begin{equation}
\begin{split}
I(y) &= y \hspace{1mm}C(y) - \int_0^y x f(x)\hspace{1mm} \textrm{d}x, \\
     &= y\hspace{1mm} C(y) - \left\{\int_0^y \frac{\textrm{d}}{\textrm{d}x}\left[x\hspace{1mm} C(x)\right] \textrm{d}x - \int_0^y C(x)\hspace{1mm}\textrm{d}x\right\},\\
     &=\int_0^y C(x) \hspace{1mm}\textrm{d}x,\\
     &\leq y\hspace{1mm} C(y).
\end{split}
\end{equation}

\noindent Therefore, we have

\begin{equation}
\begin{split}
\mathbb{E}[\left|x-y\right|] &\leq 2\int_0^\infty f(y) \hspace{1mm} y \hspace{1mm}C(y)\hspace{1mm} \textrm{d}y,\\
                             &= 2\int_0^1 y\hspace{1mm} C \hspace{1mm}\textrm{d}C,\\
                             &=\frac{2}{\log(1-p)} \int_0^1 \log\left(1-C^{1/q}\right)C \hspace{1mm}\textrm{d}C,\\
                             &=\frac{H\!\left(2q\right)}{\left|\log(1-p)\right|},
\end{split}
\end{equation}

\noindent where $H\!\left(n\right) = \sum_{m=1}^n 1/m$ is the $n^\textrm{th}$ harmonic number. Here, $q$ is equal to the total number of elementary pairs that need to connect in each ``section'' at a given level, which is given by $2^{N_S-l}$, which arrives at Eq.~(\ref{eq:harmonic_eq}).

\bibliographystyle{apsrev4-1}
%\bibliography{bib_this.bib,Mendeley04.bib}
%merlin.mbs apsrev4-1.bst 2010-07-25 4.21a (PWD, AO, DPC) hacked
%Control: key (0)
%Control: author (72) initials jnrlst
%Control: editor formatted (1) identically to author
%Control: production of article title (-1) disabled
%Control: page (0) single
%Control: year (1) truncated
%Control: production of eprint (0) enabled
%

%\bibliography{Mendeley04.bib}

\vfill
\end{document}